\documentclass[reqno]{amsart}

\usepackage{t1enc} 
\usepackage{amsxtra}
\usepackage{feynmf}
\author{Dirk Prange}

\title{Causal Perturbation Theory and Differential Renormalization}

\newcommand{\supp}{\operatorname{supp}}
\newcommand{\singsupp}{\operatorname{sing\,supp}} 
\newcommand{\singord}{\operatorname{sing\,ord}} 
\newcommand{\scaledeg}{\operatorname{scal\,deg}} 
 
\newcommand{\const}{\mathit{const}}  
 
\newcommand{\menge}[1]{\ensuremath{\mathbb{#1}}} 
\newcommand{\vek}[1]{\ensuremath{\mathbf{#1}}} 
\newcommand{\Null}[1]{\ensuremath{{^{0}#1}}} 
\newcommand{\dif}{\ensuremath{\mathrm{d}}}

\newcommand{\Lint}{\ensuremath{\mathcal{L}_{\mathrm{int}}}} 
\newcommand{\Lcal}{\ensuremath{\mathcal{L}}} 
\newcommand{\D}{\ensuremath{\partial}} 
\newcommand{\R}{\ensuremath{\menge{R}}} 
\newcommand{\Rn}{\ensuremath{\menge{R}^{n}}} 
\newcommand{\Rnon}{\ensuremath{\menge{R}^{n}\setminus\{0\}}} 
 
\newcommand{\Rvon}{\ensuremath{\menge{R}^{4}\setminus\{0\}}} 
 
\newcommand{\Cunend}{\ensuremath{\mathcal{C}^{\infty}}} 
\newcommand{\Ecal}{\ensuremath{\mathcal{E}}} 
\newcommand{\Dcal}{\ensuremath{\mathcal{D}}} 
 
\newcommand{\Scal}{\ensuremath{\mathcal{S}}} 
 
\newcommand{\scp}[2]{\ensuremath{\left\langle #1, #2 \right\rangle}} 
 
\newcommand{\W}[2]{\ensuremath{W_{(#1;#2)}}} 
\newcommand{\TR}[2]{\ensuremath{T_{R(#1;#2)}}} 
\newcommand{\ord}[1]{\ensuremath{\natural #1 \natural}} 
 
\newcommand{\wick}[1]{\ensuremath{:\! #1 \!:}} 
\newcommand{\goes}{\overset{\lambda\rightarrow 0}{\longrightarrow}}

\newtheorem{thm}{Theorem} 
\newtheorem{lem}[thm]{Lemma} 
\newtheorem{prop}[thm]{Proposition}
\newtheorem*{prob}{Problem} 
\theoremstyle{definition} 
\newtheorem{defi}{Definition} 
\newtheorem{exa}{Example}
\newtheorem*{Woperation}{$W$-operation}  
\theoremstyle{remark} 
\newtheorem*{rem}{Remark}

\begin{document}
\begin{fmffile}{grafen}
\hfill\raisebox{2cm}[0cm][0cm]{\sffamily\large  DESY 97-211}

\maketitle
\vspace{-1cm}
\begin{center}
  \sffamily
  \small
  II. Institut f\"ur Theoretische Physik\\
  Universit\"at Hamburg\\
  Luruper Chaussee 149\\
  D-22761 Hamburg, Germany\\
  \verb+dirk.prange@desy.de+
\end{center}

\begin{abstract} 
  In Causal Perturbation Theory the process of renormalization is 
  precisely equivalent to the extension of time ordered distributions to 
  coincident points. This is achieved by a modified Taylor 
  subtraction on the corresponding test functions. I show that the 
  pullback of this operation to the distributions yields expressions 
  known from Differential Renormalization. The subtraction is equivalent 
  to BPHZ subtraction in momentum space. Some examples from Euclidean 
  scalar field theory in flat and curved spacetime will be presented.
\end{abstract}

\section*{Introduction}
Calculations in perturbative QFT are performed primarily in momentum 
space. The computation of a given contribution to the \( S \)-Matrix is 
done by writing down Feynman rules and applying a certain choice of 
renormalization scheme to the resulting expression. For a reasonable 
renormalization scheme it should be proved to work to all orders and 
so produce a finite \( S \)-Matrix, as for example in the case of BPHZ 
renormalization.

But today we consider the principle of locality to be of special
importance, and hence a local formulation of perturbation theory
should exist. Indeed this was elaborated upon by Epstein and Glaser
\cite{pap:ep-gl} following earlier ideas of Bogoliubov
\cite{bk:bogol}. Their approach is called Causal Perturbation Theory
(CPT). Based on a set of axioms they constructed the \( S \)-Matrix as
a formal power series inductively. The process of renormalization
occurs only once in every step. All lower order contributions are
already renormalized. This corresponds to the determination of all
divergent subgraphs in the traditional approach and simplifies the
proof of the construction to all orders. The main concept on which CPT
is based is its formulation completely in configuration space. During
the seventies there have not been many applications of it except in
the works \cite{pap:blanchard,pap:dosch}. This may be due to the fact
that Epstein and Glaser used rigorous functional analysis, so
renormalization is defined by an appropriate subtraction on test
functions, whereas physicists are used to working with distributions
in an integral kernel representation.

Later, Scharf et al.\ applied CPT to QED (see \cite{bk:scharf} and
references therein) and to non Abelian gauge theories
\cite{pap:scharf1, pap:scharf2, pap:scharf3, pap:scharf4}. But their
perturbative calculations are still performed in momentum space. In my
opinion CPT has much potential for further applications to computation
of diagrams in configuration space.

On the other hand, a renormalization scheme called Differential
Renormalization \cite{pap:DR,pap:smir-zavDR} has gained attention. It 
is appreciated for its simplicity especially since no regularization 
procedure is needed. Differential Renormalization works in 
configuration space on a large number of examples, but a proof for all 
orders is missing.

In the present paper I will show that the integral kernel
representation of the Epstein-Glaser renormalization procedure
exactly yields Differential Renormalization. It will lead to a
simple formula for the computation of diagrams. I will apply it to
some examples from Euclidean scalar field theory and use the results
for renormalization group computations. 

CPT mainly relies on the principles of causality, translation
invariance and the singularity structure of the Feynman propagator.
Brunetti and Fredenhagen \cite{prep:fred} implemented CPT on a
globally hyperbolic spacetime by giving a local generalization of
translation invariance. Here the local causality structure is
preserved and the Feynman propagator is known to have Hadamard form. I
will show how the corresponding distributions can be renormalized in 
the Euclidean case. This is achieved by an appropriate translation of 
their representations from flat spacetime to curved spacetime.

\section{The extension of distributions}
Following \cite{lect:stora, vorl:fred} it turns out that 
renormalization in CPT actually is an extension of distributions from 
the subspace of test functions whose support does not contain the 
origin to the space of all test functions. To treat the most general 
solution of that problem we will be concerned with the space of 
distributions \( \Dcal'(\Rn) \), the dual of \( \Dcal(\Rn) \), the 
space of test functions with compact support. Let \( 
\alpha=\{\alpha_{1},\dots,\alpha_{n}\}\in\menge{N}^{n} \) be a 
multi-index, we set \( |\alpha|=\sum_{i=1}^{n}\alpha_{i} \) and \( 
\alpha!=\prod_{i=1}^n\alpha_{i}! \).
\begin{equation}
\D^{\alpha}=\frac{\D^{|\alpha|}} {{\D x_{1}}^{\alpha_{1}}\cdots{\D x_{n}}^{\alpha_{n}}}
\end{equation}
is a partial differential operator of order \( |\alpha| \).
\begin{rem}
Note that all operations on distributions like differentiation and 
transformations of their arguments are defined by the 
corresponding operations on test functions.
\end{rem}
This fact is referred to as \emph{``in the sense of distributions''}. 
Writing
\begin{equation}
T(\varphi)=\int\dif^{n}x\,T(x)\varphi(x),
\quad T\in\Dcal'(\Rn),\quad\varphi\in\Dcal(\Rn), 
\label{def:intkern}
\end{equation}
we call \( T(x) \) integral kernel of \( T \). Let \( 
\Dcal(\Rnon)=\{\varphi\in\Dcal(\Rn)|0\not\in\supp(\varphi)\} \) 
denote the subspace of test functions whose support does not contain 
the origin and \( \Dcal'(\Rnon) \) its dual.%
\footnote{The existence of the extension is guaranteed by the 
Hahn-Banach theorem. A solution for homogenous distributions can be 
found in \cite{bk:hoermander}[Chap. III.2]} 
Now we state the 
\begin{prob}
Given a distribution \( \Null{T}\in\Dcal'(\Rnon) \), how can we 
construct an extension \( T\in\Dcal'(\Rn) \), such that \( 
\Null{T}(\varphi)=T(\varphi) \) for \( \varphi\in\Dcal(\Rnon) \)?
\end{prob}
The solution of this problem requires the introduction of a quantity
that measures the singularity of the distribution at the origin
\cite{bk:steinmann}.
\begin{defi}
A distribution \( T\in\Dcal'(\Rn) \) has 
scaling degree \( s \) at \( x=0 \), if
\begin{equation}
s=\inf\{s'\in\menge{R}|\lambda^{s'}T(\lambda x)
\overset{\lambda \rightarrow 0}{\longrightarrow} 0 \text{ in the sense 
of distributions} \}.
\end{equation}
Let \( s \) be denoted by \( \scaledeg(T) \), and define 
\(\singord(T):=[s]-n \), the singular order.%
\footnote{\( [s] \) is the largest integer that is smaller than or 
equal to \( s \).} 
\end{defi}
The definition also holds if \( T\in\Dcal'(\Rnon) \). Take the 
\( \delta \)-distribution as an
\begin{exa}
For \( \delta\in\Dcal'(\Rn) \) one has: \( \delta(\lambda x)= 
|\lambda|^{-n}\delta(x) \). The scaling degree of \( \delta \) is \( n 
\), the singular order is zero.
\end{exa}
The scaling degree of some special compositions of distributions can be 
computed quite easily. We state 
\begin{prop}
\label{propSkalengrad} 
Let \( T\in\Dcal'(\Rn) \) or \( \Dcal'(\Rnon) \), \( 
\scaledeg(T)=s \) and \( \beta \) be a multi-index.
\begin{enumerate}
\item \( \scaledeg(x^{\beta}T)=s-|\beta| \). \label{prop1} 
\item \( \scaledeg(\D^{\beta}T)=s+|\beta| \). \label{prop2} 
\item \( \scaledeg(w)\leq0,\scaledeg(wT)\leq s, w\in\Dcal(\Rn) \). 
\label{prop3} \item \( \scaledeg(T_{1}\otimes 
T_{2})=s_{1}+s_{2},\text{ if } 
\scaledeg(T_{i})=s_{i},\quad i=1,2 \). \label{prop4}
\end{enumerate}
\end{prop}
The proof is skipped, we only note that all statements follow directly 
from the translation of the words ``in the sense of distributions'' 
and the use of the Banach-Steinhaus theorem (principle of uniform boundedness, 
applied to distributions) on point~\ref{prop3}. 
\begin{exa}
The scaling degree of \( \delta^{(\alpha)}\in\Dcal'(\Rn) \) is \( 
|\alpha|+n \). The singular order is \( |\alpha| \).
\end{exa}
The solution of the problem depends on the sign 
of the singular order. Let us consider the simple case first.
\begin{thm}
\label{thm:fortsetzungleicht} 
Let \( ^{0}T\in\Dcal'(\Rnon) \) with scaling degree \( s<n \). 
Then there exists a unique \( T\in\Dcal'(\Rn) \) with scaling degree 
\( s \) and \( T(\varphi)={^{0}T}(\varphi) \) for all \( 
\varphi\in\Dcal(\Rnon) \).
\end{thm}
The proof can be found in \cite{prep:fred}.  If the scaling degree is
not smaller than the space dimension, the singular order \( \omega \)
is zero or positive. In that case theorem~\ref{thm:fortsetzungleicht}
guarantees a unique extension on test functions that vanish at the
origin up to order \( \omega \).  Thus a general extension can be
defined after performing a projection into that subspace. This is
achieved by a kind of modified Taylor subtraction, called the
\begin{Woperation}
Let \( \Dcal^{\omega}(\Rn) \) be the subspace of test functions 
vanishing up to order \( \omega \) at \( 0 \). Define
\begin{gather}
\W{\omega}{w}:\Dcal(\Rn)\mapsto\Dcal^{\omega}(\Rn), 
\quad\varphi\mapsto\W{\omega}{w}\varphi, \notag \\
\left(\W{\omega}{w}\varphi\right)(x)=\varphi(x)-w(x)\sum_{|\alpha|\leq\omega} 
\frac{x^{\alpha}}{\alpha!}\left(\D^{\alpha}\frac{\varphi}{w}\right)(0),
\label{def:W}
\end{gather}
with \( w\in\Dcal(\Rn), w(0)\not=0 \).
\end{Woperation}
The action of \( \W{\omega}{w}\) on \(\varphi \) can be written as 
\begin{equation}
\left(\W{\omega}{w}\varphi\right)(x)
=\sum_{|\beta|=\omega+1}x^{\beta}\varphi_{\beta}(x),
\label{def:phibeta}
\end{equation}
with \( \varphi_{\beta}\in\Dcal(\Rn) \). It has the nice property 
 \begin{equation}
 \W{\omega}{w}w\varphi=w\W{\omega}{1}\varphi.
 \label{eq:Wwphi}
 \end{equation}
With \( (\D^{\alpha}x^{\gamma})(0)=\gamma!\delta_{\alpha}^{\gamma} \) 
it follows for \( |\gamma|\leq\omega \):
\begin{equation}
\W{\omega}{w}wx^{\gamma} =w\W{\omega}{1}x^{\gamma}\equiv0.
\label{eq:Wwxgamma}
\end{equation}
Now we can discuss the general case.
\begin{thm}
\label{thmfortsetzungschwer} 
Let \( \Null{T}\in\Dcal'(\Rnon) \) with scaling degree \(s\geq n\).
Given \( w\in\Dcal(\Rn) \) with \( w(0)\not=0 \), a multi-index \(
\alpha, |\alpha|\leq\omega \) and constants \(
C^{\alpha}\in\menge{C}\), then there is exactly one distribution \(
T'\in\Dcal'(\Rn) \) with scaling degree \( s \) and following
properties:
\begin{enumerate}
\item \( \scp{T'}{\varphi}=\scp{\Null{T}}{\varphi}\quad\forall  
\varphi\in\Dcal(\Rnon) \), \label{erst}
\item \( \scp{T'}{wx^{\alpha}}=C^{\alpha} \). \label{zweit}
\end{enumerate}
\( T' \) is given by:
\begin{equation}
\scp{T'}{\varphi}=\scp{T}{\W{\omega}{w}\varphi}+\sum_{|\alpha|\leq\omega} 
\frac{C^{\alpha}}{\alpha!}\left(\D^{\alpha}\frac{\varphi}{w}\right)(0).
\label{def:T'}
\end{equation}
Here \( T \) is the unique extension by
theorem~\ref{thm:fortsetzungleicht}, \( \W{\omega}{w} \) is given by
(\ref{def:W}) and \( \omega \) is the singular order of \( \Null{T}
\).
\end{thm}
The proof can be found in \cite{prep:fred}.
We see that in the case of non-negative singular order, the extension 
is not unique. It is fixed by a finite set of complex numbers \( 
C^{\alpha} \). Let us look at the next
\begin{exa}
\label{ex:deltafhochn} 
The \( n \)-th power of the scalar Feynman propagator \( 
(i\Delta_{F})^{n}(x)=\theta(x^{0}){\Delta_{+}}^{n}(x) 
+\theta(-x^{0}){\Delta_{+}}^{n}(-x) \) is a distribution on \Rvon. We 
compute the scaling degree of \({\Delta_{+}}^{n}\).
\begin{align*}
{\Delta_{+}}^{n}(\lambda x)
&=(2\pi)^{-3n}\int\prod_{i=1}^{n}\frac{\dif^{3}\vek{p}_{i}}{2\omega_{\vek{p}_{i}}} 
e^{\sum_{i=1}^{n}(-i\omega_{\vek{p}_{i}}\lambda{x}^{0}+i\vek{p}_{i}\lambda\vek{x})} 
\\
&=\lambda^{-2n}(2\pi)^{-3n}
\int\prod_{i=1}^{n}\frac{\dif^{3}\vek{p}_{i}}{2\sqrt{(\lambda m)^{2}+{\vek{p}_{i}}^{2}}} 
e^{\sum_{i=1}^{n}(-i\sqrt{(\lambda m)^{2}+{\vek{p}_{i}}^{2}}x^{0}+i\vek{p}_{i}\vek{x})} 
\\
&=\lambda^{-2n}{\Delta_{+}}^{n}(x,\lambda m)
\\
&\goes\lambda^{-2n}{D_{+}}^{n}(x).
\end{align*}
Here \( D_{+} \) denotes the massless scalar two point function. 
Hence the scaling degree is \( 2n \). The application of the \( W 
\)-operation with \( \omega=2n-4 \) yields the extension to all 
test functions. The computation can be done similarly for the Euclidean
propagator.
\end{exa}
We turn to another example that seems to have caused some confusion in 
classical physics (see e.g.\ \cite{bk:feynman2}). 
\begin{exa}[The self energy of the electron]
In electrostatics the electric potential of an electron at the origin 
is given by the Green's function of the Laplace equation in \(3\) 
dimensions.
\begin{align*}
\Delta\phi&=-4\pi\rho=4\pi e\delta &
&\Rightarrow &
\phi&=-\frac{e}{r}\in\Dcal'(\R^{3}).
\end{align*}
The electric field is
\begin{equation*}
\vek{E}=-\vek{\nabla}\phi=-\frac{e\vek{r}}{r^{3}}\in\Dcal'(\R^{3}).
\end{equation*}
Since \( \singsupp(\vek{E})=\{0\} \) it follows:
\begin{equation*}
{\vek{E}}^{2}=\frac{e^{2}}{r^{4}}\in\Dcal'(\R^{3}\setminus\{0\}).
\end{equation*}
The singular order is \( 1 \). Hence there is an extension to all test 
functions by the \( W \)-operation. We can define the energy density 
\( U={\vek{E}}^{2} \) as the following distribution:
\begin{equation}
\scp{U}{\varphi}:=\scp{{\vek{E}}^{2}}{\W{1}{w}\varphi}.
\label{def:U}
\end{equation}
An electron at rest has self energy \( E=1/(4\pi)\scp{U}{1} \). The 
choice of \( \varphi\equiv1 \) is possible due to sufficient 
convergence at long range. The same holds for \( w \) in (\ref{def:U}). 
\begin{equation*}
E=\frac{1}{4\pi}\scp{U}{\W{1}{1}1}+C^{0}=C^{0},
\end{equation*}
as all \( C^{\alpha},\ |\alpha|=1 \) vanish. We can determine \( C^{0} 
\) by the requirement that the mass of the electron is purely 
electromagnetic, i.e.
\begin{equation*}
E=mc^{2}.
\end{equation*}
\end{exa}
In the following I will suppress the distinction between \( 
^{0}T \) and \( T \) in the case that the scaling degree is smaller 
than the space dimension. This should lead to no confusion since the 
extension is unique in that case.
\subsection{The integral kernel representation} 
To use standard vocabulary we will call the extended distribution in
theorem~\ref{thmfortsetzungschwer} a \emph{renormalization}. Next we
will work out its integral kernel. If we set all \( C^{\alpha} \) to
zero we have
\begin{defi}
Let $T\in\Dcal^{\omega'}(\Rn)$ with \( \singord(T)=\omega \). The  
integral kernel \( \TR{\omega}{w} \) of its extension is given by
\begin{equation}
\scp{\TR{\omega}{w}}{\varphi}:=\scp{T}{\W{\omega}{w}\varphi}.
\label{def:TR}
\end{equation}
\end{defi}
Furthermore, we consider a family of distributions \( T_{t} \) that depend
continuously on a real parameter \( t \). If \( K \) is a real compact
interval then \( \int_{K}\dif t\scp{T_{t}}{\varphi} \) exists as a
Riemannian integral. We define
\begin{equation}
\scp{\int_{K}\dif t\,T_{t}}{\varphi}:=\int_{K}\dif t\scp{T_{t}}{\varphi}
\label{def:Tt}
\end{equation}
in the sense of distributions. Now we have 
\begin{prop}
The integral kernel \( \left( \TR{\omega}{w}w \right) \) is given by:
\begin{equation}
\left(\TR{\omega}{w}w\right)(x)=(-)^{\omega+1}(\omega+1) 
\sum_{|\beta|=\omega+1}\D^{\beta}\frac{x^{\beta}}{\beta!} 
\int_{0}^{1}\dif t 
\frac{(1-t)^{\omega}}{t^{n+\omega+1}}
T\left(\frac{x}{t}\right)w\left(\frac{x}{t}\right).
\label{eq:TRw}
\end{equation} 
\end{prop}
\begin{proof}
The Taylor expansion of \( \varphi \) at the origin is:
\begin{gather}
\varphi(x)=\sum_{|\alpha|=0}^{\omega}
\frac{x^{\alpha}}{\alpha!}(\D^{\alpha}\varphi)(0) 
+(\omega+1)\sum_{|\beta|=\omega+1}\frac{x^{\beta}}{\beta!} 
\int_{0}^{1}\dif t(1-t)^{\omega}(\D^{\beta}\varphi)(tx).
\label{eq:taylor}
\end{gather}
Hence \( (\W{\omega}{1}\varphi)(x) \) is the Taylor rest term of order 
\( \omega+1 \). Writing (\ref{eq:TRw}) as \( 
\left(\TR{\omega}{w}w\right) =\int_{0}^{1}\dif 
t\left(\TR{\omega}{w}w\right)_{t} \) we find using (\ref{def:Tt}):
\begin{align*}
\scp{\int_{0}^{1}\dif t\bigl(\TR{\omega}{w}w\bigr)_{t}}{\varphi} 
&=(\omega+1)\int_{0}^{1}\dif t\,(1-t)^{\omega}\times \\
&\qquad\qquad\times\int\dif^{n}x 
\sum_{|\beta|=\omega+1}\frac{x^{\beta}}{\beta!} 
T(x)w(x)(\D^{\beta}\varphi)(tx) \\
&=\scp{T}{w\W{\omega}{1}\varphi} \\
&=\scp{T}{\W{\omega}{w}w\varphi} \\
&=\scp{\TR{\omega}{w}}{w\varphi} \\
&=\scp{\TR{\omega}{w}w}{\varphi},
\end{align*}
where we used equation (\ref{eq:Wwphi}).
\end{proof}
Note that the differential operator in (\ref{eq:TRw}) is a \emph{weak} 
derivative. If \( w \) had no zeros (\( \Rightarrow 
w\not\in\Dcal(\Rn)\)), the integral kernel would result from a simple 
division. Later we will encounter a well known example of this.

To achieve a similar representation for the whole distribution we put 
a restriction on the test function \( w \) in (\ref{def:W}). Let
\( w(0)=1 \) and \( (\D^{\alpha}w)(0)=0 \), for 
\(|\alpha|\leq\omega\).
\begin{rem}
This is no real loss of generality since for a given \( 
v\in\Dcal(\Rn),\ v(0)\not=0\), the projection
\begin{equation}
v\mapsto w=v\sum_{|\alpha|\leq\omega}
\frac{x^{\alpha}}{\alpha!}\left(\D^{\alpha}\frac{1}{v}\right)(0)\in\Dcal(\Rn)
\label{def:projv}
\end{equation}
yields \( (\D^{\gamma}w)(0)=\delta_{0}^{\gamma} \) for \( 
|\gamma|\leq\omega \).
\end{rem}
Then \( (1-w) \) vanishes up to order \( \omega \) at \( 0 \):
\begin{align}
\W{\omega}{w}(1-w)\varphi&=(1-w)\varphi,
\label{eq:W(1-w)phi} \\
\scp{(1-w)\TR{\omega}{w}}{\varphi}&=\scp{(1-w)T}{\varphi}.
\label{eq:TR(1-w)}
\end{align}
Now the integral kernel is given by 
\begin{lem}
\label{lem:integralkern} 
With the above restrictions on \( w\in\Dcal(\Rn) \), the 
renormalized distribution \( \TR{\omega}{w} \) has the 
following integral kernel:
\begin{multline}\label{eq:TR}
\TR{\omega}{w}(x)
        =(-)^{\omega}(\omega+1) 
        \sum_{|\beta|=\omega+1}\D^{\beta}\frac{x^{\beta}}{\beta!} 
\biggl[-\int_{0}^{1}\dif t \frac{(1-t)^{\omega}}{t^{n+\omega+1}} 
T\left(\frac{x}{t}\right)w\left(\frac{x}{t}\right)+ \\
+\int_{1}^{\infty}\dif t \frac{(1-t)^{\omega}}{t^{n+\omega+1}} 
T\left(\frac{x}{t}\right)(1-w)\left(\frac{x}{t}\right)\biggr].
\end{multline}
\end{lem}
\begin{proof}
The first term of of (\ref{eq:TR}) is the integral kernel of \( 
w\TR{\omega}{w} \). A simple computation shows that the 
second term smeared with \( \varphi \) yields \( \scp{T}{(1-w)\varphi} 
\). Equation (\ref{eq:TR(1-w)}) completes the proof.
\end{proof}
If \( n=4m \), we can write
\begin{gather}
k!\sum_{|\beta|=k}\D^{\beta}\frac{x^{\beta}}{\beta!}= 
\ord{\left(\sum_{|\beta|=1}\D^{\beta}x^{\beta}\right)^{k}}= 
\ord{(\D_{\mu_{1}}x^{\mu_{1}}+\dots+\D_{\mu_{m}}x^{\mu_{m}})^{k}},
\label{eq:difinv} \\
\text{with } x=\begin{pmatrix} x^{\mu_{1}}\\ \vdots \\
x^{\mu_{m}}\end{pmatrix}\in\menge{R}^{4m}. \notag 
\end{gather}
Here \( \ord{\dots} \) denotes the ordering of differential operators 
to the left of the coordinates.

To perform some computations with formula (\ref{eq:TR}) it would be
desirable to abandon the requirement of \( w \) being a test function.
Let \( w_{m}\in\Dcal(\Rn) \) be a sequence with \(
\lim_{m\rightarrow\infty} w_{m} =: w\in\Dcal'(\Rn) \). If \(
\lim_{m\rightarrow\infty}\scp{\TR{\omega}{w_{m}}}{\varphi}\in\menge{C}
\) exists \( \forall\varphi\in\Dcal(\Rn) \), we will allow \( w \) to 
be used in the renormalization procedure. Let us consider \( 
T\in{\Dcal^{\omega}}'(\Rn) \) with singular order \( \omega \) and \( 
\singsupp(T)=\{0\} \). Choose $w(x)=\theta(1/M-|x|)=:\theta^{<}(x), 
M\in\R,M>0$, where \( |\cdot| \) denotes the Euclidean norm. Since \( 
\singsupp(T)\cap\singsupp(\theta^{<})=\emptyset \) and \( \theta^{<} 
\) has compact support, the pointwise product \( 
\theta^{<}T\in{\Ecal^{\omega}}'(\Rn)\subset{\Dcal^{\omega}}'(\Rn) \) 
exists.%
\footnote{$\Ecal=\Cunend$ and $\Ecal'$ is the space of distributions with
 compact support.}
Applying (\ref{eq:TR}) yields
\begin{equation}
\begin{split}
\TR{\omega}{\theta^{<}}(x) 
&=(-)^{\omega}(\omega+1)\sum_{|\beta|=\omega+1} 
\D^{\beta}\frac{x^{\beta}}{\beta!}\int_{1}^{M|x|}\dif t 
\frac{(1-t)^{\omega}}{t^{n+\omega+1}}T\left(\frac{x}{t}\right) \\
&=:T_{R}^{M}(x), \\
\end{split}
\label{def:TRM}
\end{equation}
For an arbitrary choice of \( w \), a scale \( M \) has to be 
introduced for dimensional reasons. This allows writing \( w(Mx) \) as 
a function of a dimensionless argument. The dependence of the 
counterterms on the scale can easily be computed:
\begin{gather}
M\frac{\D}{\D M}\scp{\TR{\omega}{w(Mx)}}{\varphi}
=\sum_{|\alpha|\leq\omega}B^{\alpha}\scp{\delta^{(\alpha)}}{\varphi}, 
\\
\text{with } B^{\alpha}=\frac{(-)^{|\alpha|+1}}{\alpha!} 
\scp{T}{Mx^{\mu}(\D_{\mu}w)(Mx)x^{\alpha}}.
\end{gather}
For \( w=\theta^{<} \) we get
\begin{equation}
B^{\alpha}=\frac{(-)^{|\alpha|}}{\alpha!} 
\scp{T}{\delta\left(\frac{1}{M}-|x|\right)x^{\alpha}}.
\label{eq:counterM}
\end{equation} 
\subsection{Momentum space and BPHZ renormalization}
Since the Fourier transformation is a map \( \Dcal\mapsto\Scal \) we 
have to restrict our distributional space to \(\Scal'\subset\Dcal'\).%
\footnote{Let \( \Scal \) be the space of \Cunend\ functions of rapid 
decrease and \( \Scal' \) its dual. We use the convention
\begin{equation*}
\widehat{\psi}(p)=\int\dif^n x\,\psi(x) e^{ipx}.  
\end{equation*}
} 
We remind the reader that the Fourier transformation is defined in the 
sense of distributions, i.e.\ \( 
\widehat{T}(\varphi):=T(\widehat{\varphi}) \).

First we start with the definition of the moments of a test function:
\begin{equation}
K^{\alpha}(\psi):=\int\dif^{n}x\,x^{\alpha}\psi(x),\quad\psi\in\Scal(\Rn).
\end{equation}
Let \( \Scal_{\omega}(\Rn) 
:=\left\{\psi\in\Scal(\Rn),K^{\alpha}(\psi)=0,|\alpha|\leq\omega\right\} 
\) be the subspace of test function with vanishing moments up to order 
\( \omega \), then it follows: 
\(\psi\in\Scal^{\omega}\Rightarrow\check{\psi}\in\Scal_{\omega}\). 
Choosing \( w \) as in Lemma~\ref{lem:integralkern} we have:
\begin{align}
K^{\alpha}(\widehat{w})&=0 \text{ for }0<|\alpha|\leq\omega, 
&K^{0}(\widehat{w})&=(2\pi)^{n},
\end{align}
and by a simple computation:
\begin{equation}
K^{\gamma}(\D^{\alpha}\widehat{w})
=(-)^{|\gamma|}\gamma!\,\delta_{\alpha}^{\gamma}\,
(2\pi)^{n},\quad\gamma\leq\alpha.
\label{eq:KvonDw}
\end{equation}
The Fourier transformation of \( W\varphi \) is:
\begin{align}
\left(\W{\omega}{w}\varphi\right)\spcheck(p) 
&=\check{\varphi}(p)-\frac{1}{(2\pi)^{n}}\sum_{|\alpha|\leq\omega} 
\frac{(-)^{|\alpha|}}{\alpha!}(\widehat{wx^{\alpha}})(-p)
\scp{\delta^{(\alpha)}}{\varphi} 
\\
&=\check{\varphi}(p)-\frac{1}{(2\pi)^{n}}\sum_{|\alpha|\leq\omega} 
\frac{K^{\alpha}(\check{\varphi})}{\alpha!}(\D^{\alpha}\widehat{w})(-p).
\end{align}
Using (\ref{eq:KvonDw}) we get:
\begin{equation}
K^{\gamma}\left(\left(\W{\omega}{w}\varphi\right)\spcheck\right)=0,
\quad|\gamma|\leq\omega,
\end{equation}
so \( {\W{\omega}{w}}\spcheck \) actually is a projector \( 
\Scal\mapsto\Scal_{\omega} \). With
\begin{equation}
\scp{\TR{\omega}{w}}{\varphi}= 
\scp{\widehat{\TR{\omega}{w}}}{\check{\varphi}}= 
\scp{\widehat{T}}{\left(\W{\omega}{w}\varphi\right)\spcheck}
\end{equation}
the integral kernel is given by
\begin{gather}
\widehat{\TR{\omega}{w}}(k)= \widehat{T}(k)- 
\sum_{|\alpha|=0}^{\omega}\frac{k^{\alpha}}{\alpha!}(\D^{\alpha}\widehat{Tw})(0).
\end{gather}
This subtraction should be understood in the sense of distributions, 
i.e.\ the subtraction on the test functions has to occur before 
smearing out. It looks similar to BPHZ renormalization which is a 
Taylor subtraction at arbitrary momentum \( q \). We will compute the 
corresponding \( w \). Let
\begin{equation}
\widehat{T_{R}^{q}}(k):= \widehat{T}(k)- 
\sum_{|\alpha|=0}^{\omega}\frac{(k-q)^{\alpha}}{\alpha!}(\D^{\alpha}\widehat{T})(q)
\label{eq:TRqk}
\end{equation}
denote the BPHZ renormalized distribution in momentum space. 
Using the Taylor rest expression similar to (\ref{eq:taylor}) and 
performing Fourier transformation we get:
\begin{align}
T_{R}^{q}(x)&=(-)^{\omega+1}(\omega+1)e^{-iqx} 
\sum_{|\beta|=\omega+1}\D^{\beta}\frac{x^{\beta}}{\beta!} 
\int_{0}^{1}\dif t \frac{(1-t)^{\omega}}{t^{n+\omega+1}}
T\left(\frac{x}{t}\right)e^{i\frac{qx}{t}}, \\
\intertext{and comparing to (\ref{eq:TRw}),}
&=\TR{\omega}{e^{iqx}}.
\label{eq:TRqx}
\end{align}
Here, the equivalence of the subtraction procedure in 
Epstein-Glaser and BPHZ renormalization can be seen explicitly. Moreover this 
cannot be achieved by the requirement \( w\in\Scal(\Rn) \). 
\section{Causal perturbation theory}
In the following I will give a very brief summary of CPT. A complete
description can be found in \cite{pap:ep-gl,bk:scharf}. 

We start with the \( S \)-Matrix as a formal power series
\begin{equation}
S(g)=1+\sum_{n=1}^{\infty}\frac{(-i)^{n}}{n!} 
\int\dif^{4}x_{1}\cdots\dif^{4}x_{n}\,T_{n}(x_{1},\dots,x_{n})
g(x_{1})\cdots g(x_{n}).
\label{def:smatrix}
\end{equation}
It is an operator valued functional. The function \( 
g\in\Dcal(\menge{R}^{4}) \) plays the role of a coupling ``constant''. 
The \( T_{n} \) are operator valued distributions in Fock space. They 
are called time ordered functions and involve free fields only. 
Epstein and Glaser stated six axioms from which the time ordered 
functions can be computed recursively. Here we cite only the most 
important one called the causality axiom:
\begin{equation}
T_{n}(x_{1},\dots,x_{n})=T_{k}(x_{1},\dots,x_{k})\,T_{n-k}(x_{k+1},\dots,x_{n}),
\label{ax:causal}
\end{equation} 
if all points \( x_{k+1},\dots,x_{n} \) are not in the causal past of
\( x_{1},\dots,x_{k} \). The inductive construction starts with \(
T_{1}=\Lint \). We assume that \(T_{n'}(x_{1},\dots,x_{n'})\) for all
\( n'<n \) exist as a sum of products of a symmetric translation
invariant numerical distribution and a Wick polynomial of fields. The
scaling degree of the numerical distribution is known at coincident  
points. Now \( T_{n} \) can be constructed up to the total diagonal \( 
x_{1}=\dots=x_{n} \) by the causality axiom. Wick's theorem ensures 
the required form. The numerical distributions have an extension to 
the diagonal which is the origin in \( \menge{R}^{4n-4} \) because of 
translation invariance. The Wick polynomials are already defined as 
operator valued distributions on the whole space. Let me emphasize 
that the translation invariance of the numerical distributions plays a 
crucial role in the whole construction.

\section{Applications}
I will give some examples from \( \phi^{4} \)-theory in lowest 
order. The Lagrangian for the self interacting scalar field is
\begin{equation}
\Lcal(x)=\frac{1}{2}\wick{\D_{\mu}\phi(x)\D^{\mu}\phi(x)} 
-\frac{m^{2}}{2}\wick{\phi^{2}(x)}-\frac{\lambda}{4!}\wick{\phi^{4}(x)}.
\label{eq:L}
\end{equation}
Now \( T_{1} \) is given by the interaction term.
\begin{equation}
T_{1}(x)=-\frac{\lambda}{4!}\wick{\phi^{4}(x)}.
\label{eq:T1}
\end{equation}
Causality (\ref{ax:causal}) implies \( T_{2} \) for non coincident
points.
\begin{align}
T_{2}(x_{1},x_{2}) 
&=
\begin{cases}
T_{1}(x_{1})T_{1}(x_{2}),\quad&\text{ if } x_{1}^{0}>x_{2}^{0}, \\
T_{1}(x_{2})T_{1}(x_{1}),&\text{ if } x_{2}^{0}>x_{1}^{0},
\end{cases} 
\notag \\
&=\frac{\lambda^{2}}{(4!)^{2}}\Bigl[\wick{\phi^{4}(x_{1})\phi^{4}(x_{2})}+
\label{disc} \\
&\quad+16\Delta_{F}(x_{1}-x_{2})\wick{\phi^{3}(x_{1})\phi^{3}(x_{2})}+ 
\label{kaefer} \\
&\quad+72\,(i\Delta_{F})^{2}(x_{1}-x_{2})\wick{\phi^{2}(x_{1})\phi^{2}(x_{2})}+ 
\label{fisch} \\
&\quad+96\,(i\Delta_{F})^{3}(x_{1}-x_{2})\wick{\phi(x_{1})\phi(x_{2})}+ 
\label{setsun} \\
&\quad+24\,(i\Delta_{F})^{4}(x_{1}-x_{2}) \Bigr].
\label{vakuum}
\end{align}

To give some explicit results for the numerical distributions we 
will now turn to their corresponding Euclidean counterparts. The 
singular support of these distributions is the origin only. Hence we 
can use (\ref{def:TRM}) for the renormalization.

\subsection{The massless theory}
\label{subsec:masselos}
The Green's function of the Laplace equation in four dimensions is
\begin{equation}
D_{F}(x)=\frac{1}{4\pi^{2}}\frac{1}{x^{2}}.
\label{def:DF0}
\end{equation}
It has singular order $-2$. Consider the contribution to the two 
point resp.\ four point vertex function.
\subsubsection{The one-loop graph}
The ``fish'' graph (\ref{fisch}) has singular order zero, so we get 
\begin{equation}
\left.{D_{F}}^{2}\right|_{R}^{M}(x) 
=\frac{1}{{(2\pi)}^{4}}\frac{1}{2}\D_{\mu}x^{\mu}
\frac{\ln(M^{2}x^{2})}{(x^2)^{2}}.
\label{eq:fischrenmassenull}
\end{equation}
Since 
\marginpar{%
\begin{fmfgraph}(25,20) \fmfkeep{fisch} \fmfpen{thick}
\fmfleft{i1,o1} \fmfright{i2,o2}
\fmf{fermion}{i1,v1,o1} \fmf{fermion}{i2,v2,o2} 
\fmf{fermion,left}{v1,v2} \fmf{fermion,left}{v2,v1} \fmfdot{v1,v2}
\end{fmfgraph}%
}
all extensions only differ by a \( \delta \)-term we have
\begin{align}
M\frac{\D}{\D M}\left.{D_{F}}^{2}\right|_{R}^{M}(x) 
&=\frac{1}{8\pi^{2}}\delta(x)& \ &\text{ or }&
\D_{\mu}\frac{x^{\mu}}{(x^2)^{2}}&=2\pi^{2}\delta(x),
\label{eq:countertermfisch}
\end{align}
by (\ref{eq:counterM}). This also could have been seen by expressing
\(\frac{x^{\mu}}{(x^2)^{2}}=-\frac{1}{2}\D^{\mu}\frac{1}{x^{2}}\),
which is unique by theorem~\ref{thm:fortsetzungleicht}.
\subsubsection{The two-loop graph}
The singular order of the ``setting sun'' (\ref{setsun}) is 2.  Hence
we have
\begin{equation}
\left.{D_{F}}^{3}\right|_{R}^{M}(x) 
=\frac{1}{{(2\pi)}^{6}}\frac{1}{4}
  \D_{\mu}\D_{\nu}\D_{\sigma}x^{\mu}x^{\nu}x^{\sigma}   
  \frac{1}{(x^2)^{3}}\left[\ln(M^{2}x^{2})+M^{2}x^{2}-4M\sqrt{x^{2}}+3\right].
\label{eq:m0setsun}
\end{equation}
Terms two, 
\marginpar{%
\begin{fmfgraph}(25,20) \fmfkeep{setsun} \fmfpen{thick}
  \fmfleft{i} \fmfright{o} \fmf{fermion}{i,v1,v2,o} \fmffreeze
  \fmf{fermion,left}{v1,v2} \fmf{fermion,left}{v2,v1} \fmfdot{v1,v2}
\end{fmfgraph}%
}
three and four have singular order 0, 1 and 2. As they are
zero outside the origin, they must be proportional to \( \delta \) and
its first resp.\ second derivatives. As there is no Euclidean
invariant combination of \( \D \) and \( \delta \), term two has to be
zero. Using (\ref{eq:counterM}) we get by comparison with \(
M\frac{\D}{\D M} \) on (\ref{eq:m0setsun}):
\begin{align}
\D_{\mu}\D_{\nu}\D_{\sigma}x^{\mu}x^{\nu}x^{\sigma}     
\frac{1}{(x^2)^{3}}&=\frac{\pi^{2}}{2}\Delta\delta(x), \label{eq:counter1}\\
\D_{\mu}\D_{\nu}\D_{\sigma}x^{\mu}x^{\nu}x^{\sigma}
\frac{1}{(x^2)^{2}}&=4\pi^{2}\delta(x), \label{eq:counter2} \\
\D_{\mu}\D_{\nu}\D_{\sigma}x^{\mu}x^{\nu}x^{\sigma}
\frac{1}{{\sqrt{x^{2}}}^{5}}&\equiv0, \label{eq:counter3}
\end{align}
where \( \Delta \) is the Laplacian. Now we will turn to 
\subsection{The massive theory}
\label{subsec:masse} 
The Green's function of the Euclidean Klein-Gordon equation is
\begin{equation}
\Delta_{F}(x)
=\frac{1}{{(2\pi)}^{2}}\frac{mK_{1}\left(m\sqrt{x^{2}}\right)}{\sqrt{x^{2}}}.
\label{def:DF}
\end{equation}
As $K_{1}(x)\propto 1/x$ for $x\rightarrow 0$, the singular order 
of $\Delta_{F}=-2$. We will compute 
\subsubsection{The one-loop graph}
With \( \singord({\Delta_{F}}^{2})=0 \) and (\ref{def:TRM}) we get: 
\begin{multline}
\left.({\Delta_{F}}^{2})\right|_{R}^{M}(x) 
=\frac{1}{32\pi^{4}}\D_{\mu}x^{\mu} 
\biggl\{\frac{m^{2}}{x^{2}}\left[{K_{1}}^{2}\left(m\sqrt{x^{2}}\right) 
-K_{0}\left(m\sqrt{x^{2}}\right)K_{2}\left(m\sqrt{x^{2}}\right)\right]+%
\\
\qquad-\frac{m^{2}}{M^{2}}\frac{1}{(x^2)^{2}}\left[{K_{1}}^{2}\left(\frac{m}{M}\right) 
-K_{0}\left(\frac{m}{M}\right)K_{2}\left(\frac{m}{M}\right)\right]\biggr\}.%
\label{eq:fischren}
\end{multline}
By 
\marginpar{\fmfreuse{fisch}}
writing \( x^{\mu}\dots=\D^{\mu}\dots \) which is unique we can
express it as:
\begin{multline}  
=\frac{m^{2}}{{2(2\pi)}^{4}}
\Delta\biggl\{{K_{0}}^{2}\left(m\sqrt{x^{2}}\right)-{K_{1}}^{2}\left(m\sqrt{x^{2}}\right) 
+\frac{K_{0}\left(m\sqrt{x^{2}}\right)K_{1}\left(m\sqrt{x^{2}}\right)}{m\sqrt{x^{2}}}+ 
\\
+\frac{1}{2M^{2}x^{2}}\left[{K_{1}}^{2}\left(\frac{m}{M}\right) 
-K_{0}\left(\frac{m}{M}\right)K_{2}\left(\frac{m}{M}\right)\right]\biggr\}.
\label{eq:DRfischm}
\end{multline}
This can be compared to the corresponding expression in \cite{pap:DRmass}.
\subsubsection{The two-loop graph}
Applying formula (\ref{def:TRM}) leads to an integral \( \int\dif 
s\,(s-\const)^{2}{K_{1}}^{3}(s) \) that is difficult to solve. But if 
we use the expansion
\begin{equation}
\frac{m^{3}{K_{1}}^{3}(mr)}{r^{3}}
=\frac{1}{r^{6}}+\frac{3m^{2}}{4r^{4}}\left(\ln\left(\frac{m^{2}r^{2}}{4}\right) 
+\ln(\gamma^{2})-1\right)+R(r^{-2}),
\end{equation}
we can renormalize the first and second summand with singular order \(
2,0 \) respectively. With
\begin{equation}
\left.\frac{\ln\left(\frac{m^{2}x^{2}}{4}\right)}{(x^2)^{2}}\right|_{R}^{M} 
=\frac{1}{4}\D_{\mu}x^{\mu}
\frac{\ln\left(\frac{m^{4}x^{2}}{16M^{2}}\right)\ln(M^{2}x^{2})}{(x^2)^{2}}
\end{equation}
and the previous results from subsection~\ref{subsec:masselos} we get:
\begin{multline}
\left.({\Delta_{F}}^{3})\right|_{\mathit{MR}}^{M}(x) 
=\frac{1}{{(2\pi)}^{6}}\biggl\{ 
\frac{1}{4}\D_{\mu}\D_{\nu}\D_{\sigma}x^{\mu}x^{\nu}x^{\sigma} 
\frac{1}{(x^2)^{3}}\left[\ln(M^{2}x^{2})+M^{2}x^{2}+3\right]+\\
+\frac{3m^{2}}{16}\D_{\mu}x^{\mu}\frac{1}{(x^2)^{2}}
\left[\ln\left(\frac{m^{4}x^{2}}{16M^{2}}\right)
+2\ln(\gamma^{2})-2\right]\ln(M^{2}x^{2})+
R(x^{-2})\biggr\},
\label{eq:setsunMS}
\end{multline}
where \marginpar{\fmfreuse{setsun}} the subscript \( \mathit{MR} \)
denotes the ``minimal renormalization'', i.e.\ every summand is
renormalized with its singular order.

\subsection{The renormalization group}
The parameter \( M \) plays the role of a renormalization scale. It 
enters the theory by purely dimensional reasons as an argument of the 
``function'' \( w \). We will give the lowest order contributions to 
the \( \beta,\ \gamma \) and \( \gamma_{m} \) functions in the 
renormalization group. They can be read off by solving the 
renormalization group equation
\begin{equation}
\left[M\frac{\D}{\D M}+\beta(g)\frac{\D}{\D g}
+m\gamma_{m}(g)\frac{\D}{\D m}-n\gamma(g)\right] 
\Gamma_{n}(x_{1},\dots,x_{n})=0
\label{eq:rengroup}
\end{equation}
to order \( g^{2} \), where \( g=\lambda/(16\pi^{2}) \). Using the 
expansions
\begin{align}
\beta(g,m,M)&=M\frac{\D g}{\D M} 
=\sum_{n=2}^{\infty}\beta_{n}g^{n}, \label{def:beta} \\
\gamma_{m}(g,m,M)&=M\frac{\D\ln m}{\D M} 
=\sum_{n=2}^{\infty}\gamma_{m,n}g^{n}, \label{def:gammam} \\
\gamma(g,m,M)&=\frac{M}{2}\frac{\D\ln Z}{\D M} 
=\sum_{n=2}^{\infty}\gamma_{n}g^{n},
\label{def:gamma}
\end{align}
we find \( \beta_{2}=3 \) and \( \gamma_{2}=\frac{1}{12} \) for the
massless theory. Here we had to add a term to the setting sun
proportional to (\ref{eq:counter2}) to achieve \( \gamma_{m}\equiv0
\).

In the massive theory we find 
\begin{equation}
\beta_{2}=3\frac{m^{2}}{M^{2}}{K_{1}}^{2}\left(\frac{m}{M}\right),
\end{equation}
hence \( \beta_{2}\rightarrow3 \) if \( M\rightarrow\infty \). This 
result also holds for the corresponding minimal renormalization. In 
that scheme we get
\begin{align}
\gamma_{2}&=\frac{1}{12},\\
\gamma_{m,2}&=
\frac{2M^{2}}{3m^{2}}+\frac{1}{2}\ln\left(\frac{m^{2}}{4M^{2}}\right)
+\ln(\gamma)-\frac{5}{12},
\end{align}
by using (\ref{eq:setsunMS}). 

\subsection{Curved spacetime}
Let \(\mathcal{M}\) be a globally hyperbolic manifold with a metric
\(g\). Wick polynomials were defined in \cite{pap:fred1} using
techniques from microlocal analysis. Then CPT was implemented in
\cite{prep:fred} for scalar \( \phi^{4} \)-theory. The Feynman
propagator is known to have Hadamard structure \cite{pap:kay-wald}:
\begin{equation}
\Delta_{F}\propto 
\frac{\Delta^{\frac{1}{2}}}{2\sigma} 
+v\ln(2\sigma)+w,
\label{def:DeltaF}
\end{equation}
where \( \sigma,\Delta,\ v, \) and \( w \) are smooth functions on the 
manifold and an appropriate \( i\epsilon \) regularization has to be 
chosen. By using a chart it can be seen 
to have the same scaling degree as in flat spacetime. The function \( 
\sigma \) is half the square of the geodesic distance which is 
unique in every sufficiently small neighbourhood. Let \( 
g=\mathrm{det}(g_{ab}) \). The Van-Vleck-Morette determinant
\begin{equation}
\Delta(p,p')=-\frac{1}{\sqrt{g(p)\,g(p')}}\mathrm{det}(-\sigma_{ab'}(p,p'))
\label{def:vanvleck}
\end{equation}
fulfills the following differential equation:
\begin{equation}
\nabla_{a}(\Delta\,\sigma^{a})=4\Delta.
\label{eq:vanvleck}
\end{equation}
The vector index on \( \sigma \) denotes the covariant derivative as 
usual. If we expand
\begin{align}
v&=\sum_{n=0}^{\infty}v_{n}\sigma^{n},&
w&=\sum_{n=0}^{\infty}w_{n}\sigma^{n},
\label{eq:entw}
\end{align}
in powers of \( \sigma \), the coefficients (except \(w_{0}\)) are 
determined by the Hadamard recursion relations, see
\cite{pap:dewitt}. 

Following \cite{prep:fred} the causal construction of the second order
\(S\)-Matrix is the same as in flat spacetime. The only step left is 
to perform the extension to coincident points.

Therefore we turn to the corresponding Euclidean distributions, so we
can transfer our results from the previous examples.%
\footnote{For the $\sigma$ calculus see e.g. \cite{bk:fulling}}
They may be compared with \cite{pap:luescher, pap:bunch3}. Again we 
only use the minimal renormalization scheme. 
\subsubsection{The one-loop graph}
This is given by 
\begin{multline}
\left.{\Delta_{F}}^{2}\right|_{MR}^{M} \\
=\frac{1}{16\pi^{4}}\left(\left.\frac{\Delta}{(2\sigma)^{2}}\right|_{R}^{M}+ 
2\frac{\Delta^{\frac{1}{2}}v\ln(2\sigma)}{2\sigma}+ 
2\frac{\Delta^{\frac{1}{2}}w}{2\sigma}+2vw\ln(2\sigma)
+v^{2}\ln^{2}(2\sigma)+w^{2}\right).
\label{eq:deltafquadrat}
\end{multline}
Only \marginpar{\fmfreuse{fisch}} the first term needs renormalization
and we get
\begin{equation}
\left.\frac{\Delta}{(2\sigma)^{2}}\right|_{R}^{M}
=\frac{1}{2}\nabla_{a}\sigma^{a}\frac{\Delta\ln(2M^{2}\sigma)}{(2\sigma)^{2}}.
\label{eq:fischrencst}
\end{equation}
\subsubsection{The two-loop graph}
Here we use the expansion (\ref{eq:entw}) to determine the terms that 
have to be renormalized. Then we have  
\begin{align}
\left.{\Delta_{F}}^{3}\right|_{MR}^{M} =\frac{1}{(2\pi)^{6}} 
\Biggl(&\left.\frac{\Delta^{\frac{3}{2}}}{(2\sigma)^{3}}\right|_{R}^{M}+ 
3\left.\frac{\Delta v_{0}\ln(2\sigma)}{(2\sigma)^{2}}\right|_{R}^{M}+ 
3\left.\frac{\Delta w_{0}}{(2\sigma)^{2}}\right|_{R}^{M}+ 
3\frac{\Delta \bar{v}\ln(2\sigma)}{2\sigma}+ \notag \\
&+3\frac{\Delta \bar{w}}{2\sigma} 
+3\frac{\Delta^{\frac{1}{2}}v^{2}\ln^{2}(2\sigma)}{2\sigma}
+6\frac{\Delta^{\frac{1}{2}}vw\ln(2\sigma)}{2\sigma} 
+3\frac{\Delta^{\frac{1}{2}}w^{2}}{2\sigma}+ \notag \\
&+v^{3}\ln^{3}(2\sigma)+3v^{2}w\ln^{2}(2\sigma)+2vw^{2}\ln(\sigma)+w^{3}\Biggr),
\label{eq:setsunrencst}
\end{align}
with \marginpar{\fmfreuse{setsun}}
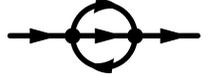
\begin{align}
\bar{v}&=\frac{1}{2}\sum_{n=0}^{\infty}v_{n+1}\sigma^{n}, &
\bar{w}&=\frac{1}{2}\sum_{n=0}^{\infty}w_{n+1}\sigma^{n}. 
\end{align}
The first term is found to be
\begin{equation} 
\left.\frac{\Delta^{\frac{3}{2}}}{(2\sigma)^{3}}\right|_{R}^{M} 
=\frac{1}{4}\bigl(\nabla_{a}\nabla_{b}\nabla_{c}\Delta^{\frac{1}{2}} 
-3\nabla_{a}\nabla_{b}{\Delta^{\frac{1}{2}}}_{;c} 
+3\nabla_{a}{\Delta^{\frac{1}{2}}}_{;bc} 
-{\Delta^{\frac{1}{2}}}_{;abc} \bigr) 
\sigma^{a}\sigma^{b}\sigma^{c}\frac{\ln(2M^{2}\sigma)}{(2\sigma)^{3}}.
\label{eq:setsunrencst1}
\end{equation}
Similarly the second and third terms can be computed:
\begin{align}
\left.\frac{\Delta v_{0}\ln(2\sigma)}{(2\sigma)^{2}}\right|_{R}^{M}
&=\frac{1}{4}\bigl(\nabla_{a}v_{0}-{v_{0}}_{;a}\bigr)\sigma^{a} 
\frac{\Delta\ln(2M^{2}\sigma)\ln\left(\frac{2\sigma}{M^{2}}\right)}{(2\sigma)^{2}}, 
\\
\left.\frac{\Delta w_{0}}{(2\sigma)^{2}}\right|_{R}^{M}
&=\frac{1}{2}\bigl(\nabla_{a}w_{0}-{w_{0}}_{;a}\bigr)\sigma^{a}
\frac{\Delta\ln(2M^{2}\sigma)}{(2\sigma)^{2}}.
\end{align}
Without further knowledge about the Feynman 
propagator, the \( \beta \)-function can be evaluated to lowest 
order, leading to the result \( \beta_{2}=3 \) in this renormalization 
scheme. The calculation requires the use of the identity
$\nabla_{a}\frac{\Delta\sigma^{a}}{(2\sigma)^{2}} 
=2\pi^{2}\delta(p,p')$. 

\section*{Conclusions}
The elegant method of CPT is suited not only for the investigation of
renormalizability but also for the performance of perturbative
computations. The subtraction procedure on the test functions can be
pulled back to the distributions yielding Differential
Renormalization. Therefore it is possible to work in the standard
integral kernel representation.

As the whole procedure is formulated in configuration space, it can be
transfered to distributions on a manifold. This enables one to give
compact expressions for the renormalization of quantum fields in
curved spacetime, at least in the Euclidean case.

\section*{Acknowledgements}
I am grateful to K. Fredenhagen for many helpful discussions and to 
Marek Radzikowski and  Klaus Bresser for their comments on the manuscript.

\bibliography{mrabbrev,literatur}
\end{fmffile}
\end{document}